\documentclass[twocolumn,showpacs,preprintnumbers,amsmath,amssymb,prb,superscriptaddress]{revtex4}

\usepackage{graphicx}

\newcommand{\cumu}[2]{\langle\!\langle #1 \rangle\!\rangle_{#2}}

\newcommand{\C}{\mathcal{C}}

\begin{document}
\title{Charge transfer statistics and entanglement in normal-quantum dot-superconductor hybrid structures}
\author{H. Soller and A. Komnik}
\affiliation{Institut f\"ur Theoretische Physik,
Ruprecht-Karls-Universit\"at Heidelberg,\\
 Philosophenweg 19, D-69120 Heidelberg, Germany}
\date{\today}

\begin{abstract}
We analyze the full counting statistics (FCS) of a single-site
quantum dot coupled to multiple metallic electrodes in the normal
state and a superconductor for arbitrary transmission. We present
an analytical solution of the problem taking into account the full
energy dependence of the transmission coefficient. We
identify two transport processes as sources of entanglement
between the current carriers by observing positive cross current
correlations. Furthermore, we consider ferromagnetic electrodes
and show how they can be used as detectors in experiments
violating the Bell-Clauser-Horne-Shimony-Holt inequality.
\end{abstract}
\maketitle
Correlations represent one of the main ingredients of quantum
mechanics.
They may persist even if the particles are causally disconnected
which allows for experimental Bell inequality tests, quantum
cryptography and quantum teleportation \cite{nielsen}. In quantum
optics using e.~g. parametric
downconversion \cite{PhysRevLett.61.2921} many interesting
experiments investigating the fundamentals of quantum mechanics
became possible \cite{zei2}. Solid state entanglers have only
recently been shown to work experimentally using a superconducting
finger coupled to normal leads directly \cite{2009arXiv0910.5558W,PhysRevLett.95.027002,PhysRevLett.97.237003} or via two quantum
dots \cite{PhysRevLett.104.026801,19829377}. They rely on the
non-local or crossed Andreev reflection (CAR) in multi-terminal
structures with a superconducting lead where an incident electron
from one of the electrodes is reflected by the superconductor into
another electrode. This process produces entanglement between the
current carriers in the two leads since the correlations of the
spin singlet (Cooper pair) are transferred to the spatially
separated electrons. The nature of CAR has been investigated
further using ferromagnetic
contacts \cite{beckmann,PhysRevLett.93.197003}.\\
Recently, the nonlocal conductance in superconductor hybrid
structures has been extensively studied theoretically  using
different approaches \cite{0295-5075-54-2-255,PhysRevB.70.174509,PhysRevLett.74.3305.2,deutscher:487,PhysRevB.68.174504,PhysRevB.76.224506,PhysRevB.74.214512,PhysRevB.63.165314}. On the one hand the conductance properties of
interacting quantum dots with a superconducting lead have been
discussed \cite{yeyati-2007-3,PhysRevB.63.094515,PhysRevLett.104.026801,Braggio2011155}. On the other hand the full counting
statistics (FCS) of multi-terminal systems with a chaotic cavity
and a superconducting connector were investigated
\cite{boerlin-2002-88,morten-2006-74,morten-2007-89,morten-2008-81,morten-2008-78}. Among other things the
cross-correlation noise has been found to be an efficient tool for
discrimination of the current caused by CAR from other current
contributions \cite{0295-5075-67-1-110,PhysRevLett.89.046601,PhysRevB.61.8125,PhysRevB.68.214501,PhysRevB.53.16390,PhysRevB.65.134522,torres,bouchiat-2003-14,lesovik-2001-24}.\\

Even in its basic realization the noninteracting quantum dot
usually shows a highly nonlinear transmission leading to
nonlinear current-voltage relations. Combined with a
superconducting electrode, the single-particle density of states
of which is itself highly nonlinear due to the finite gap, the
resulting structure is expected to possess transport properties
with a nontrivial voltage dependence. Thus far these aspects have
not been fully taken into account.
In this paper we would like to close this gap by considering a
quantum dot modelled by a noninteracting resonant level. This
setup enables one to access the full energy dependence of
transmission characteristics generated by the superconducting
density of states (DOS) as
well as by the energy-dependent transmission of the dot.\\

The quantity of our primary interest is the FCS in terms of the
cumulant generating function (CGF) $\chi$. It represents a very
convenient tool for the calculation of a variety of transport
properties. It is directly related to the probability distribution
$P(Q)$ to transfer $Q$ elementary charges during a fixed very long
measurement time $\tau$. By a simple derivation with respect to
some parameters (counting fields) $\chi$ gives all cumulants
(irreducible momenta) of $P(Q)$ \cite{levitov-1996-37,nazarov-2003-35}. The relevance of higher order cumulants has been
recently demonstrated experimentally \cite{reulet-2003-91,epub12308}. Moreover, the analytic structure of the CGF provides
deep insights into the different transport processes
\cite{levitov-2004-70,PhysRevB.50.3982,PhysRevB.63.201302,PhysRevLett.91.187001,PhysRevLett.98.056603}. As we shall
demonstrate later, in the case of the multi-terminal quantum dot
considered here, the FCS will reveal the different facets of
entanglement that we expect to be observable in upcoming
experiments.

\begin{figure}[ht]
\includegraphics[width=7cm]{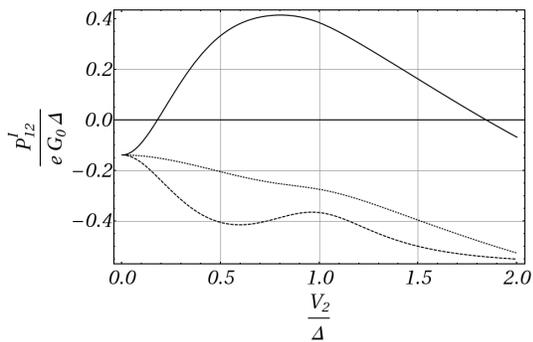}
\caption{Cross Correlations $P_{12}^{I}$ calculated from the CGF in equation (\ref{cgf}) with the parameters $\Gamma_s = \Delta/2 = \Gamma_1,\; \Gamma_2 = 0.2 \Delta, \;\Delta_d = 0, \; T=0.1 \Delta$ as a function of $V_2$ for $V_1 = -V_2$ (solid line), $V_1 = 0$ (dotted line) and $V_1 = V_2$ (dashed line). The three cuts show that one observes negative cross correlation for $V_1 = V_2$ and $V_1=0$ whereas a positive cross correlation maximum for $V_1 = - V_2$ close to the superconducting gap caused by AET is obtained.}
\label{fig1}
\end{figure}

The Hamiltonian for the system under consideration is given by
\begin{eqnarray}
H = H_0 + H_d + H_T + H_s.
\end{eqnarray}
The term $H_0$ describes normal or ferromagnetic electrodes in the
language of the respective electron field operators $\Psi_{\alpha
\sigma}(x)$, where $\alpha = 1, \dots, N$ numbers the electrodes.
The local density of states $\rho_{0\alpha}$ in the leads is
assumed to be very weakly energy dependent in the relevant range
of energies. The normal electrodes are modelled by noninteracting
fermionic continua held at the chemical potentials $\mu_\alpha$.
Ferromagnetic electrodes are described by the Stoner model with an
exchange energy $h_{{\rm ex}}$ as in \cite{melin-2004-39}
\begin{eqnarray*}
H_{{\rm Stoner},\alpha} &=& \sum_{k, \sigma} \epsilon_k \Psi_{k
\alpha \sigma}^+
\Psi_{k \alpha \sigma} \nonumber\\
&&  - h_{{\rm ex},\alpha} \sum_k (\Psi_{k \alpha \uparrow}^+
\Psi_{k \alpha \uparrow} - \Psi_{k \alpha \downarrow}^+
\Psi_{k\alpha\downarrow}).
\end{eqnarray*}
Consequently they can be described as fermionic continua with a
spin-dependent DOS $\rho_{0\sigma\alpha} = \rho_{0\alpha} (1+
\sigma P_\alpha)$, where $P_\alpha$ is the
polarization \cite{melin-2004-39}. The superconducting electrode is
described by the BCS Hamiltonian with the gap $\Delta$ of the
superconducting terminal \cite{PhysRev.108.1175}
\begin{eqnarray*}
H_s &=& \sum_{k,\sigma} \epsilon_k \Psi_{ks\sigma}^+ \Psi_{ks\sigma} \nonumber\\
&& +\Delta \sum_k (\Psi_{ks\uparrow}^+ \Psi_{-ks\downarrow}^+ + \Psi_{-ks\downarrow}
\Psi_{k s\uparrow}).
\end{eqnarray*}
The superconductor is kept in equilibrium as in previous
treatments of similar problems \cite{PhysRevB.50.3982}. The applied bias voltages are given by
$V_\alpha = \mu_s - \mu_\alpha = - \mu_\alpha$ (we use units where
$e = \hbar = k_B = 1$). The electron exchange between the dot and
the electrodes is given by \cite{PhysRevLett.8.316}
\begin{eqnarray*}
H_T &=& \sum_{\alpha, \sigma} \gamma_\alpha \left[d_\sigma^+ \Psi_{\alpha \sigma}
(x=0) + h.c.\right] \nonumber\\
&& + \sum_\sigma \gamma_s \left[d_\sigma^+ \Psi_{s\sigma}(x=0) + h.c.\right],
\end{eqnarray*}
where $\gamma_\alpha$ or $\gamma_s$ are the tunneling amplitudes between the dot
and the normal or superconducting electrodes. The tunneling is assumed to be local
and to occur at $x=0$ in the coordinate system of the respective electrode.
$d_\sigma$ is the annihilation operator of an electron with spin $\sigma$ on the dot.
The quantum dot in the presence of a magnetic field $B$ is modelled by $H_d =
\sum_\sigma (\Delta_d + \sigma h/2) d_\sigma^+ d_\sigma =: \sum_\sigma \delta_\sigma
d_\sigma^+ d_\sigma$, where $\Delta_d$ is the bare dot energy and in SI-units
$h=\mu_B gB$ with Bohr's magneton $\mu_B$ and the gyromagnetic ratio $g$
\cite{PhysRevB.75.235105}.\\
In order to determine the FCS we calculate the CGF $\ln
\chi({\boldsymbol \lambda}) = \ln \langle e^{i \lambda Q}\rangle$,
which depends on the counting fields ${\boldsymbol \lambda} =
(\lambda_1, \dots, \lambda_N, \lambda_s)$ in the respective
electrodes. It provides all the higher statistical moments (or
irreducible cumulants) $\cumu{Q^n}{}$ of $P(Q)$. The CGF is
calculated using the Keldysh Green's function
approach \cite{nazarov,levitov-1996-37} adapted to quantum
impurity problems in \cite{PhysRevB.73.195301,bagrets-2006-54,PhysRevB.76.241307}. According to the
generalized Keldysh approach one obtains for the
CGF \cite{levitov-2004-70}
\begin{eqnarray*}
\chi({\boldsymbol \lambda}) = \langle T_{\C} \exp \left[-i \int_{\C}
dt H_T^{\lambda(t)}\right]\rangle,
\end{eqnarray*}
where the dependence on the counting fields is contained in
\begin{eqnarray*}
H_T^{\lambda(t)} &=& \sum_{\alpha, \sigma} \gamma_\alpha \left[e^{i \lambda_\alpha(t)/2}
 d_\sigma^+ \Psi_{\alpha \sigma}(x=0) + h.c.\right] \nonumber\\
&& + \sum_\sigma \gamma_s \left[e^{i \lambda_s(t) /2} d_\sigma^+ \Psi_{s\sigma}(x=0)
+ h.c.\right].
\end{eqnarray*}
The different counting fields are nonzero only during the
measurement time $\tau$ and have different signs on the forward
and backward path of the Keldysh contour. The scattering matrix
has two energy regimes \cite{schwab-1999-59} with energy-dependent
transmission coefficients. The respective Fermi distributions of
the individual terminals are abbreviated by $n_\alpha$ and
$n_{\alpha +} := 1- n_\alpha(-\omega)$ for hole-like
contributions. This allows to express the CGF in the form given in equation (\ref{cgf}).
 \begin{widetext}
\begin{eqnarray}
\ln \chi({\boldsymbol \lambda}) &=& \frac{\tau}{\pi} \int
d\omega \left[\theta\left(\frac{|\omega| - \Delta}{\Delta}\right)
\left(\sum_\sigma \ln \left\{1 + \sum_{i,j = 1,\dots, N,s , \; i
\neq j} T_{ij\sigma}(\omega) n_i(1-n_j) (e^{i(\lambda_i
- \lambda_j)}-1) \right\} \right)\right. \nonumber\\
&& + \frac{1}{2} \theta\left(\frac{\Delta - |\omega|}{\Delta}\right) \left(\sum_\sigma \ln
\left\{\left[1+ \sum_{i,j = 1, \dots, N, \; i \neq j} T_{Aij\sigma
e} n_i(1-n_j)(e^{i(\lambda_i
- \lambda_j)}-1)\right] \right. \right. \nonumber\\
&& \times \left[1 + \sum_{i,j = 1, \dots, N, \; i \neq j}
T_{Aij\sigma h} n_{j+}(1-n_{i+})(e^{i(\lambda_i -
\lambda_j)}-1)\right] + \sum_{i = 1, \cdots, N}
 T_{Ai\sigma} \left[n_i(1-n_{i+}) (e^{2i (\lambda_i - \lambda_s)}-1)\right]\nonumber\\
&& \left. \left. \left. + \sum_{i,j = 1,\dots, N, \; i \neq j}
T_{CAij\sigma} \left[n_{j+}(1-n_i) (e^{i(2\lambda_s - \lambda_i -
\lambda_j)}-1)+ n_i(1-n_{j+}) (e^{-i(2\lambda_s - \lambda_i -
\lambda_j)}-1)\right]\right\}\right)\right],\label{cgf}
\end{eqnarray}
where we define
\begin{eqnarray}
T_{ij\sigma}(\omega) &=& 4\Gamma_i \Gamma_j (1+ \sigma P_i)(1+
\sigma P_j) /[(\omega - \delta_\sigma)^2 + (\sum_{k =1,
\dots, N} \Gamma_k
(1+ \sigma P_k) + \Gamma_{s1})^2], \; \mbox{with} \; (P_s = 0)\\
T_{Aij\sigma e} &=& 4 \Gamma_i (1+ \sigma P_i) \Gamma_j (1+ \sigma
P_j) \{(\omega -  \delta_{-\sigma})^2 + [\sum_{k = 1, \dots,
N }
\Gamma_k (1- \sigma P_k)]^2\}/ \mbox{det}_{A\sigma}(\omega),
\end{eqnarray}
\begin{eqnarray}
T_{Aij\sigma h} &=& 4\Gamma_i (1- \sigma P_i) \Gamma_j (1- \sigma
P_j) \{(\omega - \delta_\sigma)^2 + [\sum_{k = 1, \dots, N}
\Gamma_k
(1+ \sigma P_k)]^2\}/\mbox{det}_{A\sigma}(\omega),\\
T_{Ai\sigma} &=& 4\Gamma_i^2 (1- \sigma P_i) (1+ \sigma P_i) \Gamma_{s2}^2
/\mbox{det}_{A\sigma}(\omega), \;\;T_{CAij\sigma} = 4 \Gamma_i (1+ \sigma P_i)
\Gamma_j (1- \sigma P_j) \Gamma_{s2}^2 /\mbox{det}_{A\sigma}(\omega),\\
\mbox{det}_{A\sigma}(\omega) &=& (\omega - \delta_\sigma)^2
(\omega - \delta_{-\sigma})^2 + \{[\sum_{k = 1, \dots, N}
\Gamma_k (1+ \sigma P_k)]^2 + \Gamma_{s2}^2\}\nonumber \\
&& \times \{[\sum_{k = 1, \dots, N} \Gamma_k(1- \sigma P_k)]^2 + \Gamma_{s2}^2\} +
[\sum_{k=1, \dots, N} \Gamma_k(1+ \sigma P_k)]^2 (\omega - \delta_{-\sigma})^2 \nonumber \\
&& + [\sum_{k =1, \dots, N} (1- \sigma P_k)]^2 (\omega -
\delta_\sigma)^2 + 2 \Gamma_{s2}^2 (\omega - \delta_\sigma)(\omega - \delta_{-\sigma})  ,
\end{eqnarray}
\end{widetext}
The abbreviation $\Gamma_i = \pi \rho_{0i} |\gamma_i|^2/2$ is the (energy-independent) dot-lead
contact transparency with dimension energy for the normal leads. For the superconducting
leads it is affected by the energy-dependent superconducting DOS so that
\begin{eqnarray}
\Gamma_{s1} = \pi \rho_{0s} |\gamma_s|^2 |\omega|/(2 \sqrt{\omega^2 - \Delta^2})
\\
 \Gamma_{s2} = \pi \rho_{0s} |\gamma_s|^2\Delta /(2\sqrt{\Delta^2 - \omega^2}) \, .
\end{eqnarray}
For $|\omega|>\Delta$ the counting factors $e^{i (\lambda_i - \lambda_j)}$ describe
single electron transport between the different terminals. For $|\omega|< \Delta$ the
superconducting DOS only allows excitations from the Cooper pair condensate which leads
to different transport characteristics. They are described by counting factors
$e^{2i(\lambda_s - \lambda_i)}$ and $e^{2i \lambda_s - i \lambda_i - i \lambda_j}$
referring to the transfer of two particles from the superconductor to a single or two
separate terminals. The transmission coefficients $T_{Ai\sigma}$ and $T_{CAij\sigma}$
therefore refer to Andreev-reflection and CAR respectively.\\
For $\Delta \rightarrow 0$ the CGF reduces to the well-known expression for the CGF of
a noninteracting dot in a multi-terminal geometry \cite{PhysRevLett.98.056603}. The same
holds  for $\gamma_s \rightarrow 0$.\\
The FCS for a three-terminal structure at $T=0$ using a chaotic
cavity with energy-independent transmission instead of a resonant
level have been calculated
before \cite{boerlin-2002-88,morten-2008-78}. In this case the CGF
adopts a characteristic double square root form. The first square
root instead of a logarithm is due to the
diffusive transport through a chaotic cavity \cite{PhysRevLett.87.197006}. The second
square root may be explained looking at physical observables that are calculated via
derivatives of the CGF, where it leads to additional factors $1/2$. Therefore it
corresponds to the factors $1/2$ in the transparencies $\Gamma_i, \; \Gamma_{s1}$
and $\Gamma_{s2}$ that are due to the separate treatment of electrons and holes as
a consequence of the proximity effect. The transmission coefficients are different
in our case because of the energy-dependent DOS of the quantum dot. As in our case
the ones for single electron transmission, direct Andreev reflection and crossed
 Andreev reflection are proportional to $\Gamma_1\Gamma_2, \; (\Gamma_1^2 + \Gamma_2^2)
 \Gamma_s^2$
 and $2\Gamma_1 \Gamma_2 \Gamma_s^2$, respectively.\\
Considering the case of only a single normal electrode the
conductivity at low voltages and no magnetic field is given by
\begin{eqnarray*}
G_{NQS} &=& 4 \left. T_{A1\uparrow} \right|_{h=0, \; \omega = 0}, \\
&=& \frac{4e^2}{h} \left(\frac{2 \tilde{\Gamma}_1 \tilde{\Gamma}_s}{4 \Delta_d^2 + \tilde{\Gamma}_1^2 + \tilde{\Gamma}_s^2}\right).
\end{eqnarray*}
In the last step we introduced $\tilde{\Gamma}_1 = 2 \Gamma_1$,
$\tilde{\Gamma}_s = 2 \Gamma_s$ and restored SI-units. The result
coincides with the one previously obtained in \cite{PhysRevB.46.12841}.\\
The three-terminal case with two normal metal drains is of
particular interest. Positive cross correlations between the
normal electrodes can be used to probe the existence of
entanglement \cite{morten-2008-81}. Along with the noise the cross
correlation can be calculated as a second derivative of the CGF
\begin{eqnarray*}
P_{12}^{I} = - \frac{1}{\tau} \left. \frac{\partial^2 \ln
\chi({\boldsymbol \lambda}, \tau)}{\partial \lambda_1 \partial
\lambda_2}\right|_{{\boldsymbol \lambda} = 0}.
\end{eqnarray*}
Depending on the coupling and voltages choice different cross
correlations may be observed. Three different types of transport
between the normal drains are present: the direct single electron
current proportional to $\Gamma_1 \Gamma_2 (V_1 - V_2)$, the
direct Andreev reflections (DA) proportional to $2 (\Gamma_1^2 V_1
+ \Gamma_2^2 V_2) \Gamma_s^2$ or CAR proportional to $2 \Gamma_1
\Gamma_2 \Gamma_s^2 (V_1 + V_2)$. If the superconducting terminal
is weakly coupled to the quantum dot, cross correlations will
either be dominated by single electron transmission or, for $V_1
\approx V_2$, by direct Andreev reflections to a normal drain
leading to negative correlations. There are two cases in which
positive cumulants are observed. If the superconductor is coupled
better to the quantum dot than the normal terminals, positive
correlations may be observed for voltages close to the
superconducting gap and $V_1 \approx - V_2$, see Figure \ref{fig1}.
In this case CAR is strongly suppressed and one expects single
electron transmission to be dominant. However, the
energy-dependent DOS of the superconductor leads to large
transmission coefficients for double Andreev reflections from one
normal drain to the superconductor and further to the second
normal electrode.
\\
This Andreev-reflection Enhanced Transmission (AET) is also
observed if an additional broadening of the BCS DOS is taken into
account as in \cite{PhysRevB.70.174509}, where $\omega
\rightarrow \omega + i \eta_s$ with a typical experimental value
$\eta_s$ of about $10^{-2} \Delta$. AET is a robust phenomenon. In modern experiments using either InAs nanowires \cite{19829377} or carbon nanotubes \cite{PhysRevLett.104.026801} $\Gamma_s \approx \Delta$ is generically obtained. The hybridisation with the normal leads can be tuned via top gates to e.g. $\Gamma_n \approx \Gamma_s /2$. AET should then be observable via cross correlations or from the direct currents.

\begin{figure}[ht]
\includegraphics[width=7cm]{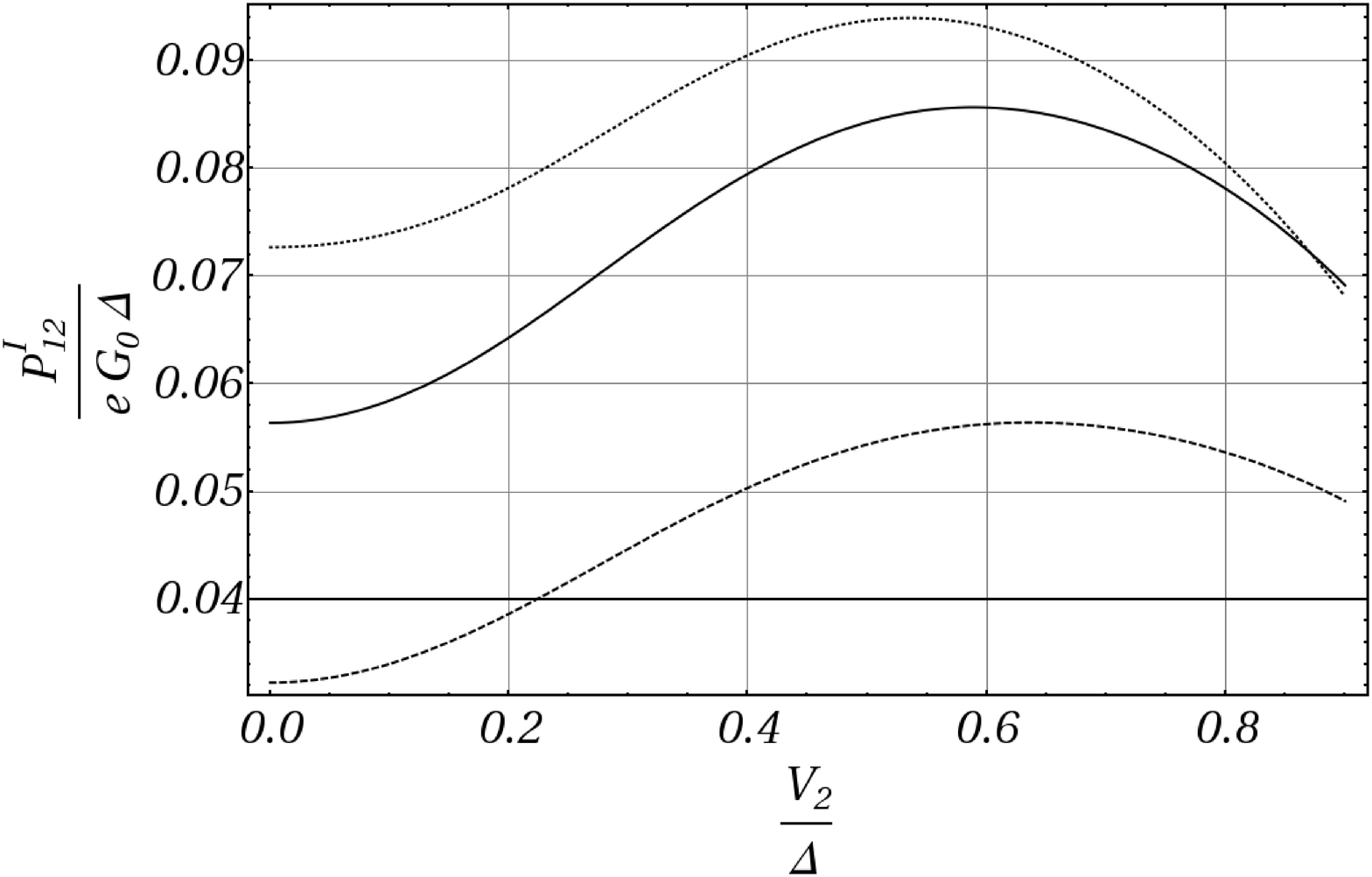}
\caption{Cross correlation $P_{12}^I$ calculated from the CGF in equation (\ref{cgf}) with the parameters $\Gamma_s = \Delta, \; \Gamma_1 = 0.4 \Delta, \; \Delta_d = 0, \; T=0.1 \Delta, \; V_1 = 0$ as a function of $V_2$. The dotted line shows the result for $\Gamma_2 = 0.05 \Delta$, the solid line is for $\Gamma_2 = 0.1 \Delta$ and the dashed line is for $\Gamma_2 = 0.15 \Delta$. Consequently the second electrode is weakly coupled. Therefore the CAR processes dominate for voltages below the gap over single electron transmission and DA leading to a positive cross correlation. This effect is enhanced for weaker coupling of the second electrode. The effect is also present at finite $V_1$ but it is weakened.}
\label{fig2}
\end{figure}

\begin{figure}[ht]
\includegraphics[width=7cm]{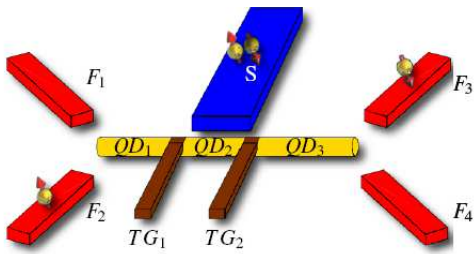}
\caption{{Schematic setup for demonstrating the violation of
Bell's inequality. A nanowire is connected to four ferromagnetic
drains $F_{1,\dots,4}$ and a superconductor $S$. The top gates
$TG_{1,2}$ may be used to divide the nanowire into three quantum
dots $QD_{1,2,3}$ with tunable coupling.}} \label{fig3}
\end{figure}

The second case for positive correlations is observed for strong
coupling of the superconductor to the quantum dot and
asymmetrically coupled normal terminals. Since the contribution by
CAR is proportional to $V_1 + V_2$ a voltage bias may be applied
via the the weakly coupled normal electrode $j=1,2$. The DA
contribution is proportional to $\Gamma_j^2$ and therefore will be
strongly suppressed for this weakly coupled drain. For voltages
well below the gap CAR dominates over single electron transmission
because a CAR is possible in two ways. In Figure \ref{fig2} we show
the cross correlation in this situation.
\\
The directions of the currents are different in the case of CAR and AET. CAR
describes electron transfer into/from the superconductor whereas in the case
of AET the superconductor only assists electron transfer between the normal
drains. The FCS formalism allows us to follow the different transport processes independently. In the case of CAR two electrons from the same Cooper pair are transferred to spatially separate electrodes. In the case of AET an electron impinging on the superconductor is retroreflected as a hole. The same Cooper pair that was generated by this Andreev reflection is then transferred further by a hole from the second normal drain that is retroreflected as an electron. In the case of CAR we therefore observe entangled electrons in the normal electrodes both in energy and spin space. Considering AET the electron and the hole originate from one Cooper pair and due to the conservation of spin and energy must also be entangled in energy and spin space. Only AET and CAR generate positive correlations which therefore indicate the presence of entanglement.\\
The asymmetric coupling necessary to observe CAR can easily be realized
experimentally by fabricating a simple NQS structure on a substrate with a
single normal drain and using an STM tip as the second weakly coupled terminal.\\
There are two further possibilities to observe CAR as the dominant transport
channel. On the one hand using additional normal drains enhances the possibilities
for CAR as can be seen directly from the CGF in equation~(\ref{cgf}). The combinatorical
factor naturally leads to a dominance of CAR. On the other hand DA and
single electron transmission can be fully suppressed by ferromagnetic drains
with antiparallel polarisation \cite{morten-2008-81}. In this case we always
observe positive correlations for voltages below the gap and not too high temperatures
if the polarisations are chosen strong enough. For $P_1 = - P_2 = 1$ we also
considered the case of three normal terminals and we obtained $P_{12}^I = 0$ in
accordance with previous calculations \cite{PhysRevLett.94.210601}.\\
A ferromagnet preferably accepts electrons of a specific spin. It may therefore
also be viewed as a detector for this specific spin direction \cite{morten-2008-81,PhysRevLett.94.210601}. This fact enables us to demonstrate a possible violation of
the Bell-Clauser-Horne-Shimony-Holt inequality along the lines of \cite{morten-2008-81,PhysRevLett.94.210601}.
First we have to include the spatial direction of magnetization into our
previous treatment, where we only considered one specific quantization axis for the
polarizations of the ferromagnet.\\
We consider a device as in Figure \ref{fig3} with four $F_n$ ($n=1,
\cdots, 4$) terminals. The drains $F_1, F_2$ and $F_3, F_4$ are
pairwise polarized in opposite directions to serve as spin
detectors. First we only consider the case with a single quantum
dot in the middle. For a simplified discussion we want to consider
the special case $P_1 = - P_2 = P_3 = - P_4$ and $\Gamma_1 =
\Gamma_2, \; \Gamma_3 = \Gamma_4$ as in
\cite{PhysRevLett.94.210601}. To avoid confusion with the
different quantization axes we limit our discussion to $h=0$. The
energy-dependent transmission coefficients for CAR in the case of
four ferromagnetic drains and $\alpha =1,2$, $\beta = 3,4$ are
given by $T_{CA\alpha \beta \sigma} = 4 \Gamma_1 \Gamma_3 (1- P_1
P_3) \Gamma_{s2}^2 /\mbox{det}_{A\sigma}(\omega)$. The
polarization in the direction-dependent case may be rewritten as
$P_i \rightarrow {\mathbf g}_i = P_i {\mathbf m}_i, \; i = 1,3$
with a unit vector ${\mathbf m}_i$ describing the magnetization
direction \cite{PhysRevLett.88.047003}. In this way we obtain the
transmission coefficients $T_{CA\alpha \beta \sigma} = 4\Gamma_1
\Gamma_3 (1- {\mathbf g}_\alpha {\mathbf g}_\beta)
\Gamma_{s2}^2/\mbox{det}_{A\sigma}(\omega)$. The differential
conductances at $T=0$ for the respective crossed Andreev
reflections to drain $\alpha$ and $\beta$ can be directly read off
via $G_{CA\alpha\beta \sigma} = \frac{4}{2\pi} \left[T_{CA\alpha
\beta \sigma}(V) + T_{CA\alpha \beta\sigma} (-V)\right]$. As CAR is a nonlocal process these conductances describe nonlocal correlations.
\\
They also give the probabilities $p_{\alpha, \beta}$
for simultaneous detection of an electron at $\alpha$ and $\beta$ by
\begin{eqnarray}
p_{\alpha, \beta} = G_{CA\alpha\beta \sigma}/\sum_{\{\gamma, \delta\}}
G_{CA\gamma \delta \sigma} \, .
\end{eqnarray}
The drains $1,2$ and $3,4$ represent spin
detectors (Alice, Bob) with respect to two possible choices for the direction
${\mathbf g}_{1,3}, \; {\mathbf g}'_{1,3}$. The spins can be detected by the
respective currents. We shall discard events with both electrons going to the
same detector by normalizing the probabilities to go to different detectors as
in \cite{PhysRevLett.94.210601}.\\
We should stress that we need to record every single event separately. To go over to a non-time-resolved measurement scheme one needs to observe not just the spin current but the spin-current fluctuations \cite{PhysRevB.66.161320}.\\
As in \cite{morten-2008-81,PhysRevLett.94.210601} we find
for the Bell parameter $\epsilon = |{\mathbf g}_1 {\mathbf g}_3 +
{\mathbf g}'_1 {\mathbf g}_3 + {\mathbf g}'_1{\mathbf g}'_3 -
{\mathbf g}_1 {\mathbf g}'_3|$. The maximum of $\epsilon$ is $2
\sqrt{2}$. Bell's inequality $\epsilon \leq 2$ is violated if the
polarization $\| {\mathbf g}_1 \| = P_1 = P_3 \geq
2^{-\frac{1}{4}}$ as in other treatments \cite{kawabata}. Such a
degree of polarisation is hard to obtain with ferromagnets but may
easily be reached using double-quantum dot
structures \cite{PhysRevB.81.075110}. This also represents a direct
measurement of the concurrence ($C$) since the maximal value of
the Bell parameter $\epsilon_{max} = 2 \sqrt{1+ C^2}$, where $C$
is a measure of entanglement with possible values between $0$ and
$1$.

The same expression for the Bell parameter has been obtained
previously by Morten et al. \cite{morten-2008-81} for a chaotic
cavity. Our result shows that the concurrence does not depend on
how the beam splitter geometry is realized. The cross correlation
however is affected by the difference in the transmission
coefficients and therefore only indicates the presence of
entanglement. The described experiment appears to be a perfect
analogue to the optical experiments using parametric
downconversion \cite{PhysRevLett.61.2921}. However, care has to be
taken with respect to the distinguishability of particles. In
optical experiments one detects spatially separated particles
where entanglement is unambiguously defined. In the case of a
single quantum dot we detect the spin direction of electrons
coming from the dot which are indistinguishable. In this situation
the state has the form of a single Slater determinant, which is
considered a nonentangled state \cite{ghirardi}. But we can easily
go over to the distinguishable case by using the full power of the
experimental setup in Figure \ref{fig3}. By tuning the top gate
voltages one can induce tunnel barriers to obtain a triple quantum
dot in series \cite{grove}. If we go over to a strongly coupled
triple dot we obtain well separated densities of states and
therefore detect distinguishable particles from $QD_1$ or $QD_3$.
In this case we can neglect resonant tunneling and evaluate the
transmission coefficients by multiplying the transmissions through
the different quantum dots. In this situation we also obtain the
same Bell parameter, however now for an unambiguously entangled
state. The correlations remain the same in the distinguishable or
indistinguishable case.

To conclude we have analytically calculated the full counting
statistics (FCS) of a noninteracting quantum dot (resonant level)
contacted by multiple metallic electrodes in the normal state as
well as by a superconducting one. The superconducting ground state
leads to positive cross correlations between the currents in the
normal drains by two transport phenomena that we identified as
crossed Andreev reflections (CAR) and Andreev-reflection enhanced
transmission (AET). Our results allow for an analysis of a
possible violation of the Bell-Clauser-Horne-Shimony-Holt
inequality that is related to a measurement of the concurrence
$C$. Interestingly, we find it to be independent of the
constellation of coupling parameters. Moreover, it coincides with
the results found for the quantum dot realization via a chaotic
cavity. Finally we discussed the question of entanglement for
indistinguishable and distinguishable particles. We have shown
that the correlations should be unaffected by this distinction and
how both cases can be addressed experimentally.

The authors would like to thank S. Maier and P. Pl\"{o}tz for many
interesting discussions. The financial support was provided by the
DFG under grant No. 2235/3, and by the Kompetenznetz "Funktionelle
Nanostrukturen III" of the Baden-W\"{u}rttemberg Stiftung (Germany).
%
\bibliography{entanglement}
%
%
%

\end{document}